\documentclass[doublecol]{epl2}

\newcommand{\avg}[1]{\left<#1\right>}

\usepackage{amsfonts}
\usepackage{amsmath}
\usepackage{graphicx}
\usepackage{amssymb}
\usepackage{color}
\usepackage{subfigure}
\usepackage[perpage]{footmisc}
\usepackage{mathtools}

\begin{document}
\title{Quantum fluctuations of the friction force induced by the dynamical Casimir emission}
\shorttitle{Quantum fluctuations of the dynamical Casimir friction force}

\author{Salvatore Butera \and Iacopo Carusotto}
\shortauthor{S. Butera and I. Carusotto}

\institute{INO-CNR BEC Center and Dipartimento di Fisica, Universit\`a di Trento, I-38123 Povo, Italy
  }

\abstract{
We study a quantum model of dynamical Casimir effect in an optical cavity enclosed by a freely moving mirror attached to a harmonic spring. The quantum fluctuations of the friction force exerted by the dynamical Casimir emission onto the moving mirror are investigated, as well as their consequences on the quantum state of the mirror oscillation. Observable signatures of the interplay of the nonlinear nature of the effective mirror-cavity coupling and of the discreteness of the emitted photons are pointed out, in particular as a fast diffusion of the mirror oscillation phase. These results are interpreted in the language of quantum field theories on curved space-times as a breakdown of the standard semiclassical theory of the backreaction.
}
\maketitle

\makeatletter
\let\toc@pre\relax
\let\toc@post\relax
\makeatother 

\section{Introduction\label{sec:Intro}}
One the most fascinating pillars of quantum mechanics is Heisenberg's indetermination principle, according to which the observable quantities of any physical system show non-trivial fluctuations even in the system's ground state. In the context of quantum field theories such as quantum electrodynamics, this implies that the {\em quantum vacuum} state is not just an empty space but crawls of virtual particles  that emerge out of the vacuum and then quickly annihilate again~\cite{Milonni-book}.  One of the most intriguing consequences of these fluctuations is the possibility of transforming the virtual photons corresponding to the zero-point fluctuations into an observable quantum vacuum radiation when the background on which the quantum field lives is modulated in either space or time~\cite{birrell1984quantum}: the related effects of dynamical Casimir emission~\cite{Moore-DCE-1970,Fulling_Davies-DCE,Davies-Fulling,Dodonov-DCE,lambrecht2005electromagnetic,KardarRMP1999} and cosmological particle creation \cite{Parker-PartCr-I,Parker-PartCr-II} have been predicted when the boundary conditions imposed to the field or the overall size of a flat space-time are varied in time; a thermal Hawking radiation~\cite{Hawking1974,Hawking1975} has been anticipated to occur in the static but strongly curved space-time around an astrophysical black hole.

So far, most of the literature on quantum field theories on curved space times~\cite{birrell1984quantum} has focused on the dynamics of the quantum field on top of a given background and, in particular, on the intensity of the quantum vacuum emission, whereas relatively fewer works have attacked the much more complex problem of the {\em backreaction} of this emission onto the background degrees of freedom of the underlying spacetime. The most celebrated such effects include the friction force exerted onto an accelerated (neutral) mirror by the dynamical Casimir photons~\cite{KardarRMP1999} or the evaporation of a black hole under the effect of Hawking radiation~\cite{fabbri2005modeling}. In the standard descriptions, the backreaction is included at a semi-classical level by plugging the expectation value of the energy-momentum tensor of the quantum field (which of course includes the emitted radiation) back into the motion equation of the background, e.g. as a source term in the Einstein equations for the space-time curvature around a black hole. More refined treatments describing the fluctuations of the energy-momentum tensor include the so-called stochastic gravity approach~\cite{Hu-StocGrav-CGQ,Hu-StocGrav-LivRev}.

In this Letter we consider a simplest yet realistic model of an optical  cavity enclosed by a freely moving mirror attached to a harmonic spring and we develop a fully quantum study of the backreaction force exerted by the dynamical Casimir emission (DCE) onto the moving mirror. Building atop our previous study of the {\em average} friction force~\cite{Butera-BR_DCE-2019}, we focus here on the {\em quantum fluctuations} of the DCE friction force around the average value and on their observable consequences. While the usual quantum theory of damping~\cite{Petruccione-book,Milburn-Walls} as well as the semiclassical theory of backreaction preserve coherent states during the dissipation process, our fully quantum theory predicts that the nonlinear form of the mirror-cavity coupling combined with the quantumness of the DCE emission are responsible for a dramatic breakdown of the semiclassical theory and give new intriguing features in the {\em quantum state} of the mechanical oscillation of the mirror.

\section{The theoretical model\label{sec:Sec1}}

\begin{figure}
\includegraphics[width=0.99\columnwidth]{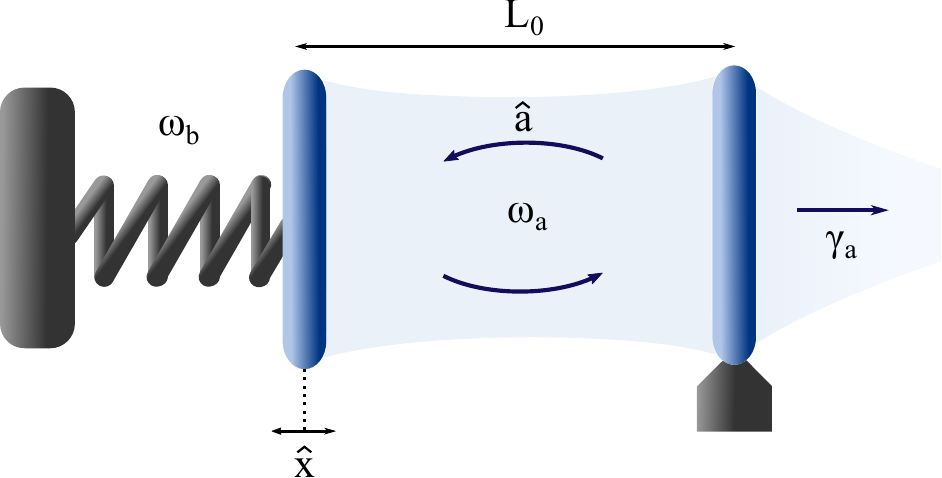}
\caption{Illustrative representation of the system under consideration. One of the cavity mirrors is allowed to harmonically oscillate around its equilibrium position and is opto-mechanically coupled to the cavity mode via the radiation pressure.}
\label{fig:sketch}
\end{figure}

We consider the system sketched in Fig.\ref{fig:sketch}, namely an optical cavity terminated on one side by a partially reflecting mirror and on the other side by a perfect mirror which is free to move under the effect of a harmonic spring of angular frequency $\omega_b$, whose equilibrium position corresponds to a cavity length $L_0$. Under a suitable quasi-resonance condition $\omega_b\approx 2 \omega_a$, we can restrict the dynamics of the cavity field to a single mode of angular frequency $\omega_a$. The radiation pressure 
is at the origin of the opto-mechanical coupling between the cavity mode and the mirror itself~\cite{Aspelmeyer_RMP}. By defining $\hat{a}/\hat{a}^\dag$ ($\hat{b}/\hat{b}^\dag$) the annihilation/creation operators of the cavity field (mechanical oscillator), the opto-mechanical coupling can be written~\cite{Law-MirFieldInt-1995} at lowest order in the mirror displacement $\hat{x}=x_{\text{ZPF}}(\hat{b}+\hat{b}^\dag)=\left(\hbar/2m\omega_b\right)^{1/2}\,(\hat{b}+\hat{b}^\dag)$ and pressure $\hat{P}=(\hbar\omega_a/2 L_0)\,\left(\hat{a}+\hat{a}^\dag\right)^2$ operators as
\begin{equation}
	\hat{H}_{\rm int}=-\hat{x}\hat{P}=-\hbar\omega_c\left(\hat{a}+\hat{a}^\dag\right)^2(\hat{b}+\hat{b}^\dag),
\label{Hint}
\end{equation}
with an optomechanical coupling strength
\begin{equation}
\omega_c=\frac{\omega_a x_{\text{ZPF}}}{2L_0}=\left(\frac{\hbar}{8m_b\omega_b} \right)^{1/2}\,\frac{\omega_a}{L_0}~.
\label{omegaC}
\end{equation}
Within a realistic~\cite{Butera-BR_DCE-2019} weak coupling regime defined by the condition $\omega_c/\omega_{a,b}\ll 1$, we can perform a rotating-wave approximation~\cite{CohenTannoudji-AtomPhot} where one neglects the anti-resonant terms in Eq.~\eqref{Hint} and only retains the resonant ones corresponding to processes where mechanical excitations of the mirror are annihilated (created) and a pair of photons is simultaneously created (annihilated) via the DCE. A more general treatment including the anti-resonant terms in the ultra-strong coupling limit $\omega_c/\omega_{a,b}\gtrsim 1$ was reported in \cite{Savasta-PRX-2018}. Within our RWA approximation, the Hamiltonian takes the simple form  
\begin{equation}
\hat{H}=\hbar\omega_a\hat{a}^\dag\hat{a}+\hbar\omega_a\hat{b}^\dag\hat{b}+\hbar\omega_c\hat{b}^\dag\hat{a}^2+\hbar\omega_c\hat{b}\left(\hat{a}^2\right)^\dag.
\label{Eq:Hamiltonian}
\end{equation}
Losses due to the coupling of the optical field to external baths of radiative and/or non-radiative origin are included at the level of the master equation for the density matrix $\hat{\rho}$, which reads
\begin{equation}
	\frac{d\hat{\rho}}{dt}=\frac{1}{i\hbar}\left[\hat{H},\hat{\rho}\right]+\mathcal{L}_{\hat{a}}[\hat{\rho}],
	\label{Eq:MasterEq}
\end{equation}
with the usual Lindblad super-operator~\cite{Petruccione-book,Milburn-Walls}
\begin{equation}
\mathcal{L}_{\hat{a}}[\hat{\rho}]\equiv\left(\gamma_a/2\right)\left(2\hat{a}\hat{\rho} \hat{a}^\dag-\hat{a}^\dag \hat{a} \hat{\rho}-\hat{\rho} \hat{a}^\dag \hat{a}\right)\,.
\label{Eq:LimbladOp}
\end{equation}
To emphasize the backreaction effects, no intrinsic mechanical damping is assumed to be acting on the mirror.

\section{Semi-classical theory\label{Sec:Semi-Clas}}
We begin our study with a comparison between the semiclassical prediction for the average force and the exact result obtained by a numerical integratiom of the master equation \eqref{Eq:MasterEq}, projected on a Fock number basis with a suitably large cut-off in the occupation numbers. In the language of the semiclassical theory of the backreaction, the role of the background is played here by the amplitude $b=\langle{\hat{b}}\rangle$ of the mirror oscillation, while the role of the energy-momentum tensor is played by the expectation values of quadratic field operators $n_a=\avg{\hat{n}_a}=\avg{\hat{a}^\dag \hat{a}}$ and $q=\avg{\hat{q}}=\avg{\hat{a}^2}$, respectively representing the population and the anomalous field correlations of the cavity mode induced by the DCE.

\begin{figure*}[htb]
\centering
\subfigure
{\includegraphics[width=5.9cm]{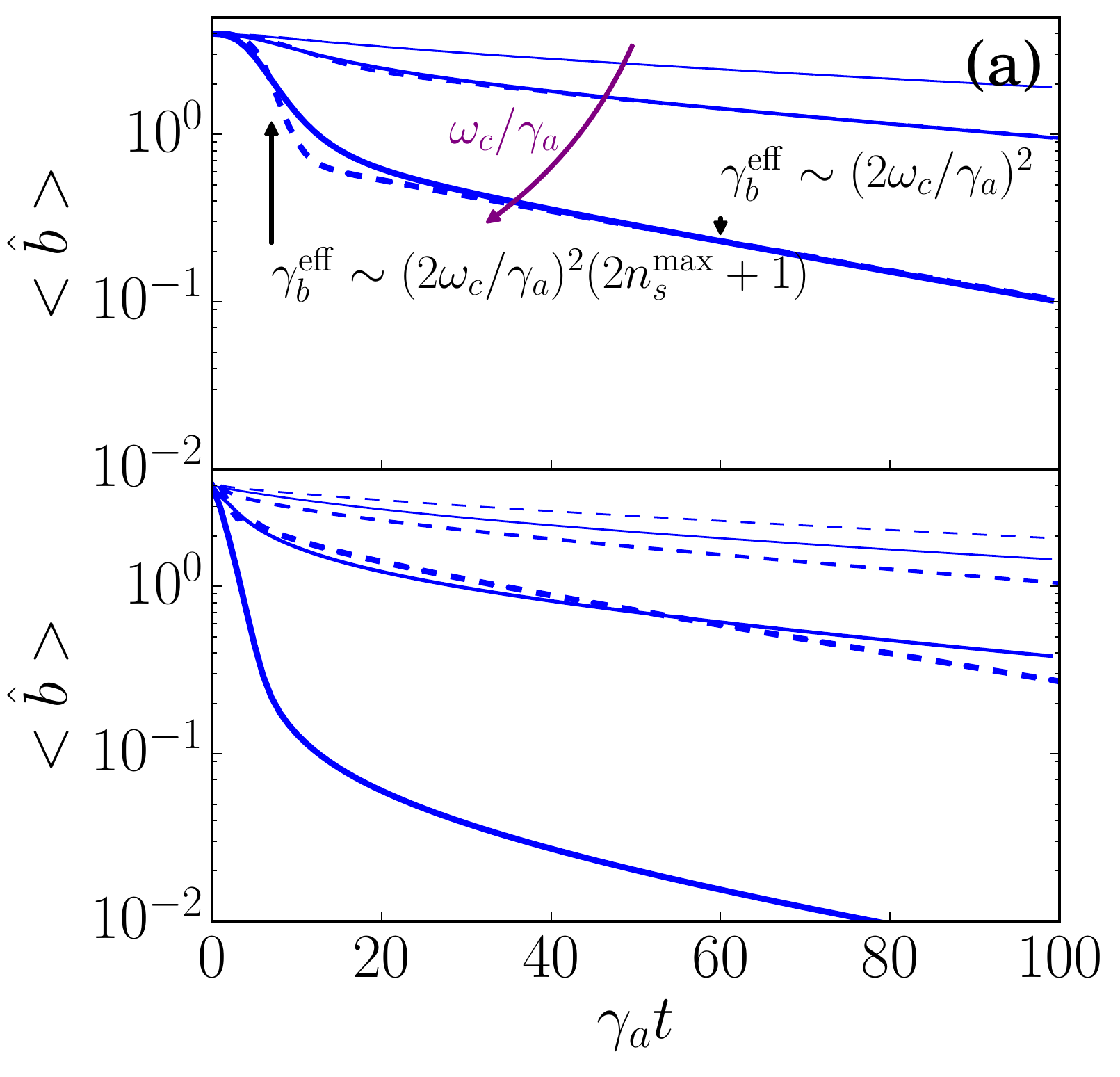}}\hfill 
\subfigure
{\includegraphics[width=5.9cm]{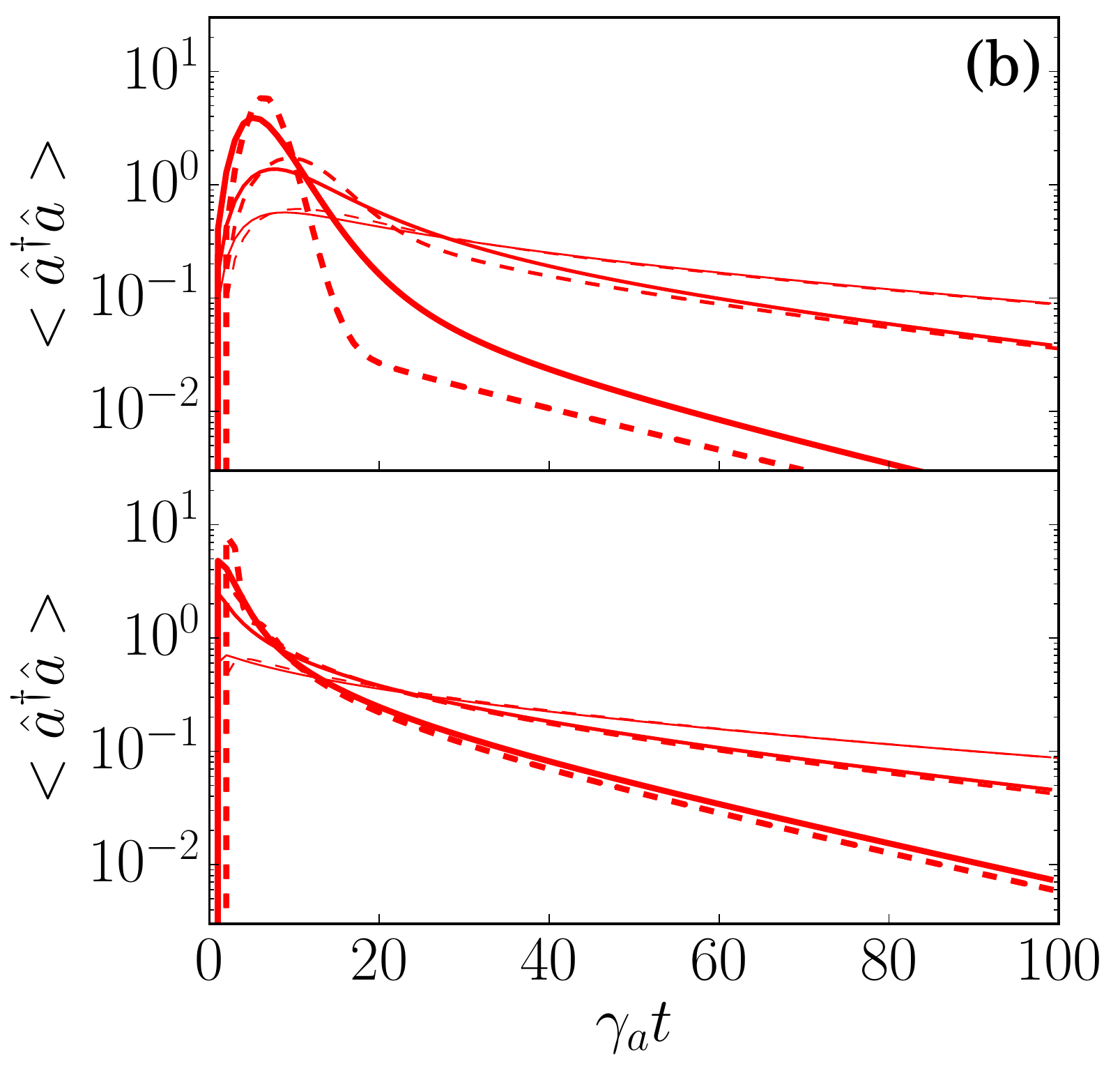}}\hfill 
\subfigure
{\includegraphics[width=5.9cm]{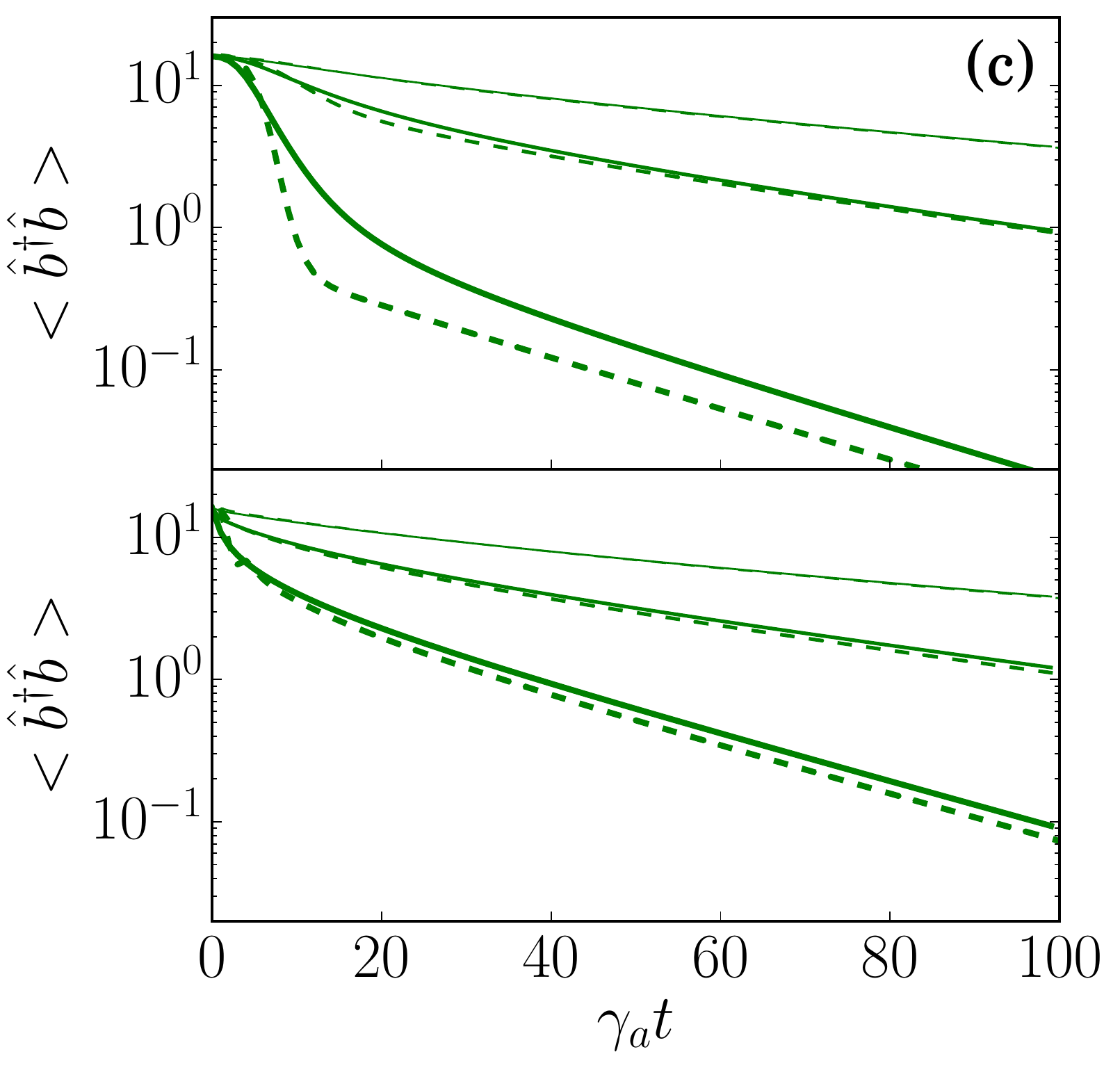}}\\
\subfigure
{\includegraphics[width=5.9cm]{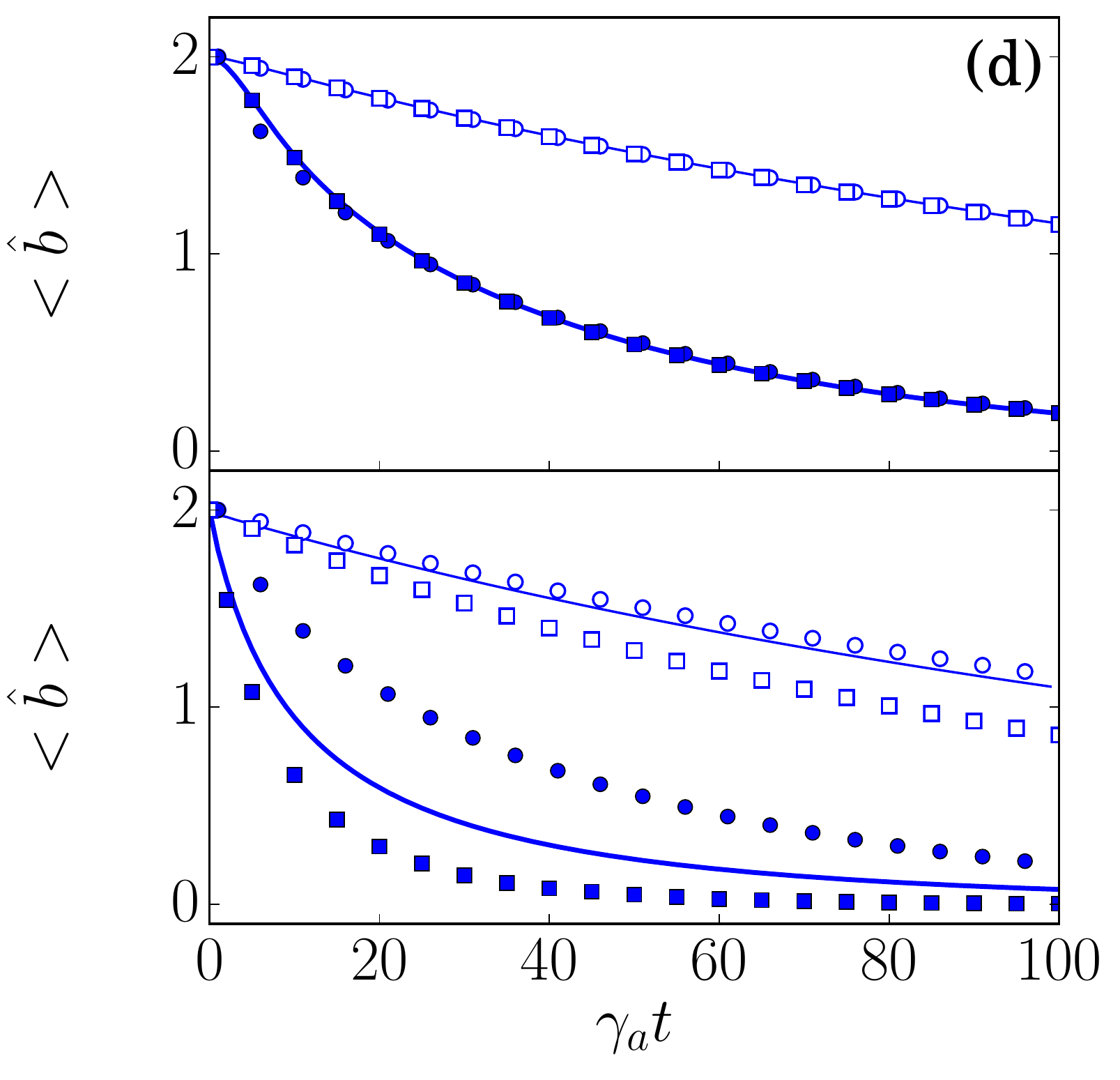}}\hfill 
\subfigure
{\includegraphics[width=6.cm]{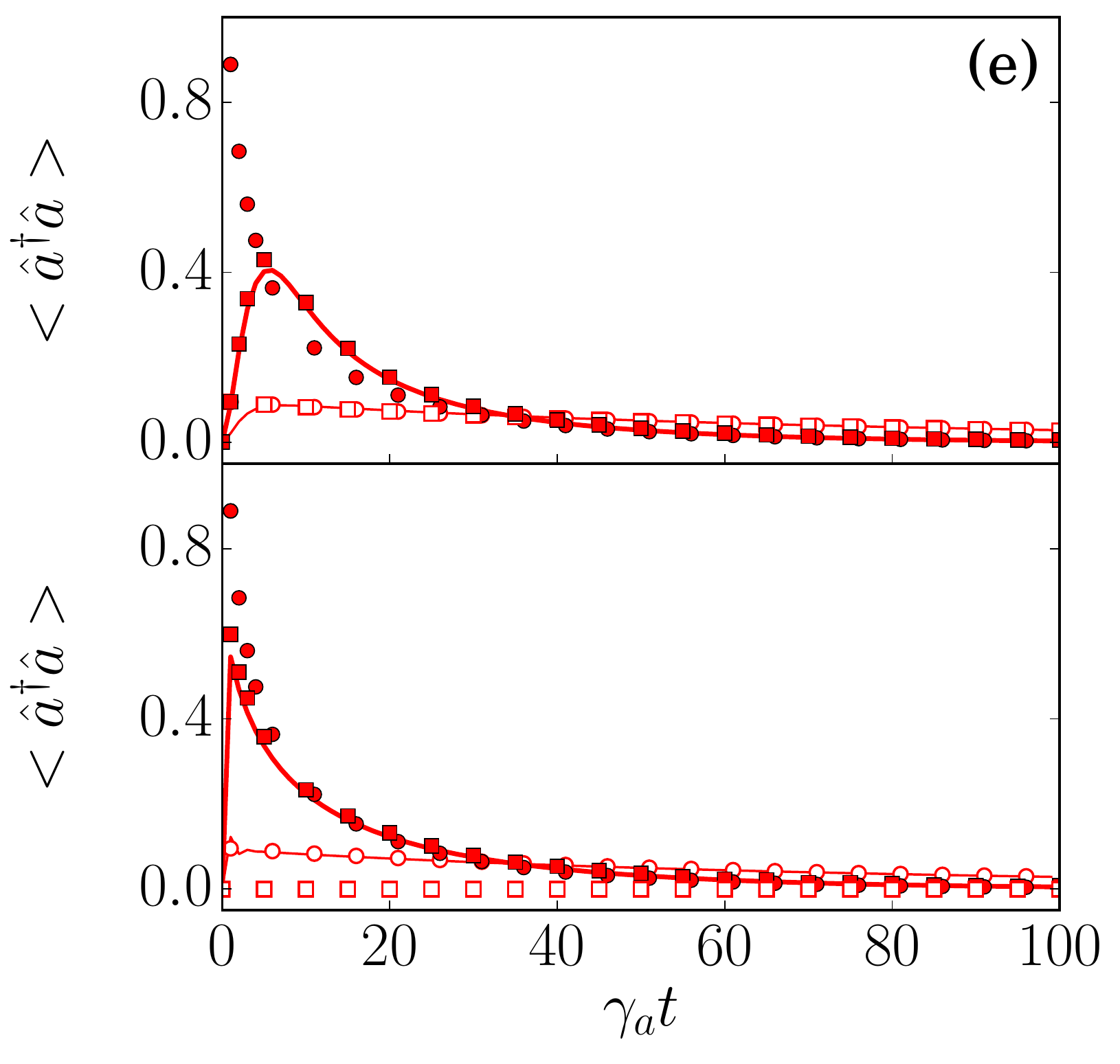}}\hfill 
\subfigure
{\includegraphics[width=5.9cm]{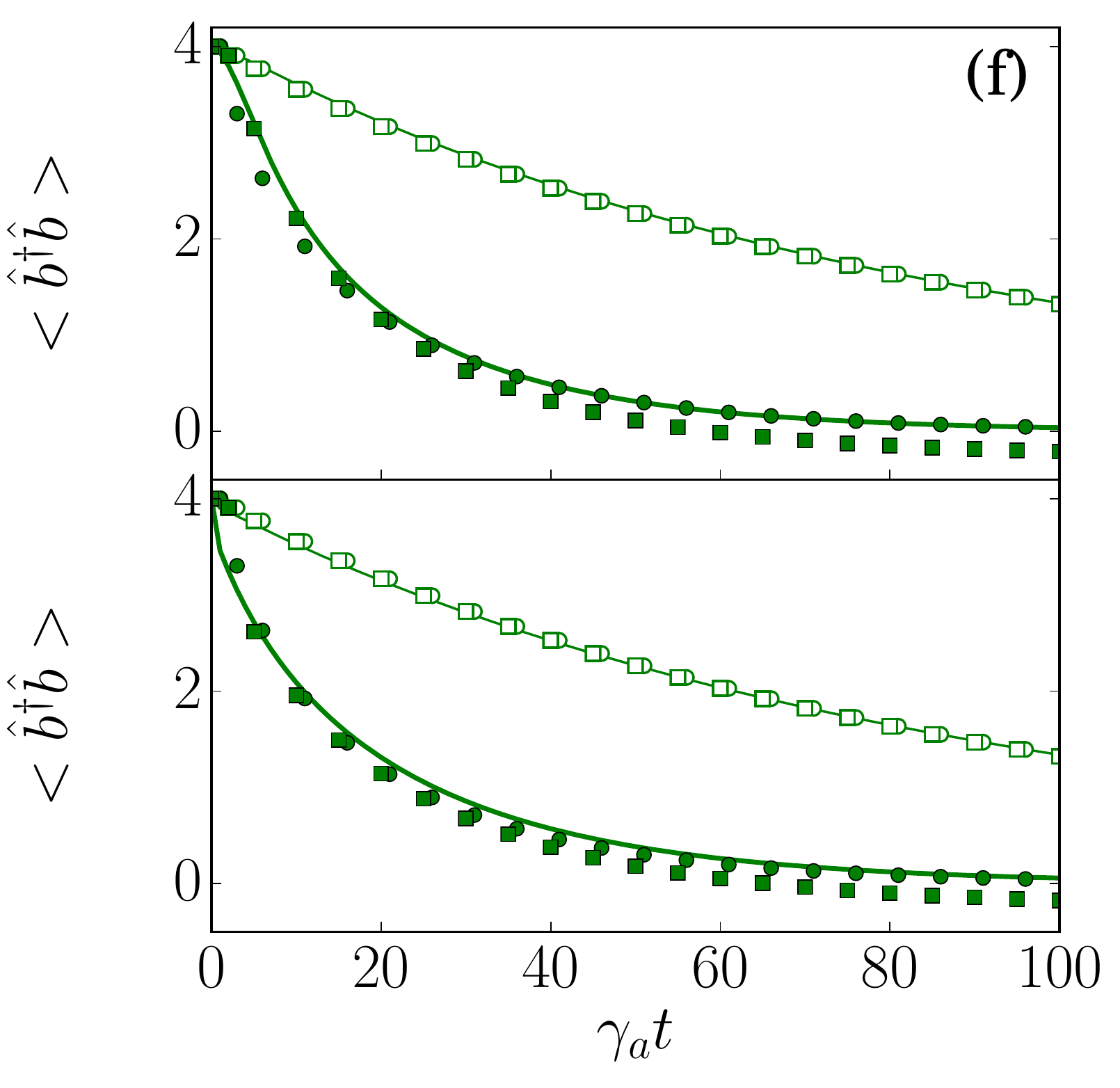}}\\ 
\caption{Time evolution of: the average oscillation amplitude of the mirror (a,d), the average number of cavity photons (b,e) and the average number of mechanical excitation quanta of the mirror (c,f). In panels (a-c), the result of the numerical integration of the full master equation \eqref{Eq:MasterEq} (solid line) is compared to the semi-classical approximation (\ref{Eq:EqMot-b}-\ref{Eq:EqMot-q}). Thicker (thinner) lines correspond to higher (lower) values of the optomechanical coupling $\omega_c/\gamma_a=1/20,\,1/15,\,1/10$. The mirror is initially prepared in a coherent state of amplitude $\langle\hat{b}\rangle=4$ and the cavity in its ground state.
In panels (d-f), the full master equation (solid line) is compared to the truncated Wigner approximation (\ref{Eq:SDE-A}-\ref{Eq:SDE-B}) (square markers) and the Born-Oppenheimer approximation \eqref{Eq:EffEqMot-b} (round markers) for coupling $\omega_c/\gamma_a=1/20$ (fine line, empty markers) and $\omega_c/\gamma_a=1/10$ (thick line, filled markers). The mirror oscillation is initially prepared in a coherent state of amplitude $\langle\hat{b}\rangle=2$ and the cavity in its ground state. In each panel, the top plot refers to the case of a resonant mirror-field interaction $\Delta=0$, while the bottom one refers to the off-resonant case $\Delta/\gamma_a=10$.
}
\label{Fig:NonLinDecay}
\end{figure*}

Working in the the rotating frame at $\omega_b$ where $q(t)=e^{-i\omega_b t}\,\bar{q}(t)$ and $b(t)= e^{-i\omega_b t}\,\bar{b}(t)$ and denoting $\Delta\equiv\omega_b-2\omega_a$ the detuning from the DCE resonance, the master equation \eqref{Eq:MasterEq} translates into the following equations of motion for the slowly varying variables:
\begin{align}
	\frac{d\bar{b}}{dt}&=-i\omega_c q,
\label{Eq:EqMot-b}\\
	\frac{dn_a}{dt}&=-\gamma_a n_a-2i\omega_c\left<\bar{q}^\dag \bar{b}\right>+2i\omega_c\left<\bar{q} \bar{b}^\dag\right>,
\label{Eq:EqMot-na}\\
		\frac{d\bar{q}}{dt}&=i\left(\Delta+i\gamma_a\right)\bar{q}-4i\omega_c\left<n_a \bar{b}\right>-2i\omega_c \bar{b}\,,
\label{Eq:EqMot-q}
\end{align}
where the cubic terms in the Hamiltonian \eqref{Eq:Hamiltonian} are responsible for the coupling to higher-order operators. 

For weak values of the coupling strength $\omega_c/\gamma_a\ll 1$, we expect that quantum fluctuations are suppressed, so that the non-factorisable component in the higher-order correlations between the field and the mirror can be safely neglected. This is the core assumption of the \emph{semi-classical} approximation. These correlators can thus be factorized as $\langle a^2 b^\dag\rangle\approx qb^*$ and $\langle a^\dag a b\rangle \approx n_a b$ (we indicate by ``$^*$'' the complex conjugate operation and we drop for simplicity the bars indicating slowly varying quantities), which leads to a non-linear set of equations that can be readily solved by numerical means. 

In Fig.~\ref{Fig:NonLinDecay}, we compare the result of this procedure (dashed lines) to the master equation \eqref{Eq:MasterEq} (solid line). At the initial time, the cavity is assumed to be in its vacuum state, while the mirror is assumed to be prepared in a classical coherent state. No external driving force is assumed to be acting on the mirror.
The panels in the left column show the evolution in time of the average amplitude $\langle \hat{b}\rangle$ of the mechanical oscillation, the one in the central column show the average number $\langle \hat{a}^\dagger \hat{a}\rangle$ of photons in the cavity mode, the ones in the right column show the average number $\langle \hat{b}^\dagger \hat{b}\rangle$ of mechanical quanta. 
The results for the $\Delta=0$ resonant case (upper plot in each panel) confirm that the semiclassical theory gets more and more accurate as the coupling $\omega_c$ is decreased. As expected, the decay gets faster for growing $\omega_c$ and a stronger acceleration is visible at short times for the largest values of $\omega_c$.

In order to get some analytical understanding of this physics, we make a further approximation step and we assume the DCE-induced relaxation dynamics of the oscillator to be much slower than the intrinsic one of the field at $\gamma_a$. This allows us to pursue a sort of \emph{Born-Oppenheimer} (BO) approximation, in which we first solve the cavity field dynamics at a fixed oscillator amplitude $b$ and then we reinsert the steady-state field correlators into the oscillator dynamics.
In this way, Eq.~\eqref{Eq:EqMot-q} directly provides a relation between the steady-state amplitudes $q_{SS}$ and $n_{a,SS}$,
\begin{equation}
	q_{SS}=\frac{2\omega_c}{\Delta+i\gamma_a}\left(2n_{a,SS}+1\right)b.
\label{Eq:Relation-Na-q}
\end{equation}
which can be substituted into Eq.~\eqref{Eq:EqMot-b} to obtain an effective evolution equation for the mirror oscillation amplitude,
\begin{equation}
	\frac{db}{dt}=-\frac{\Gamma_b(n_a)}{2} b\,.
\label{Eq:EffEqMot-b}
\end{equation}
Here the effective (complex) decay rate
\begin{equation}
	\Gamma_b(n_{a,SS})=\gamma_b^{\rm eff}\left(1+i\frac{\Delta}{\gamma_a}\right)(2n_{a,SS}+1) \label{Eq:EffGammaB}
\end{equation} 
is to be evaluated at the steady-state photon number
\begin{equation}
	n_{a,SS}=\frac{2(\gamma_b^{\rm eff}/{\gamma_a})|b|^2}{1-4(\gamma_b^{\rm eff}/{\gamma_a})|b|^2}.\label{Eq:Na-SS}
\end{equation}
obtained combining Eqs.~\eqref{Eq:EqMot-na} and \eqref{Eq:Relation-Na-q}, where the effective damping rate
\begin{equation}
\gamma_b^{\rm eff}=\frac{4\omega_c^2\gamma_a}{\gamma_a^2+\Delta^2}\,.  
\end{equation}
As expected, for a finite detuning $\Delta$ the decay rate $\rm{Re}[\Gamma_b]$ gets reduced but, at the same time, a \emph{reactive} imaginary component appears that shifts the mechanical oscillator frequency by an amount $\textrm{Im}[\Gamma_b]=[4\omega_c^2\Delta/(\gamma_a^2+\Delta^2)]\,(2n_{a,SS}+1)$. Whereas this frequency shift can be viewed as a DCE analog of the Lamb shift (which is the reactive counterpart of the spontaneous radiative decay of an atom~\cite{CohenTannoudji-AtomPhot}), it displays an additional dependence on the amplitude $b$ via the $n_a$-dependence of $\Gamma_b$.

\begin{figure}[htbp]
\centering
\subfigure[]
{\includegraphics[width=0.48\linewidth]{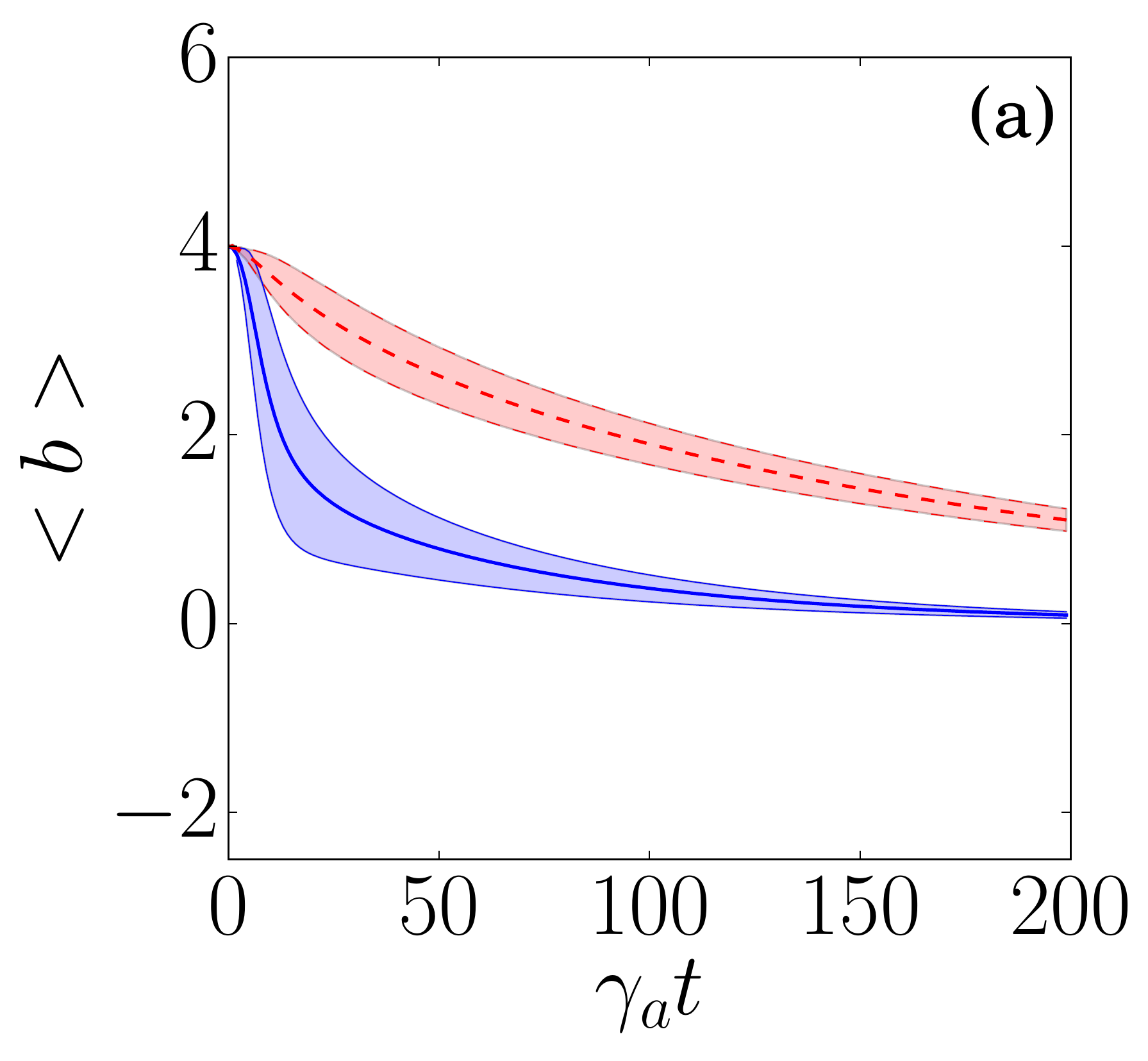}}
\subfigure[]
{\includegraphics[width=0.48\linewidth]{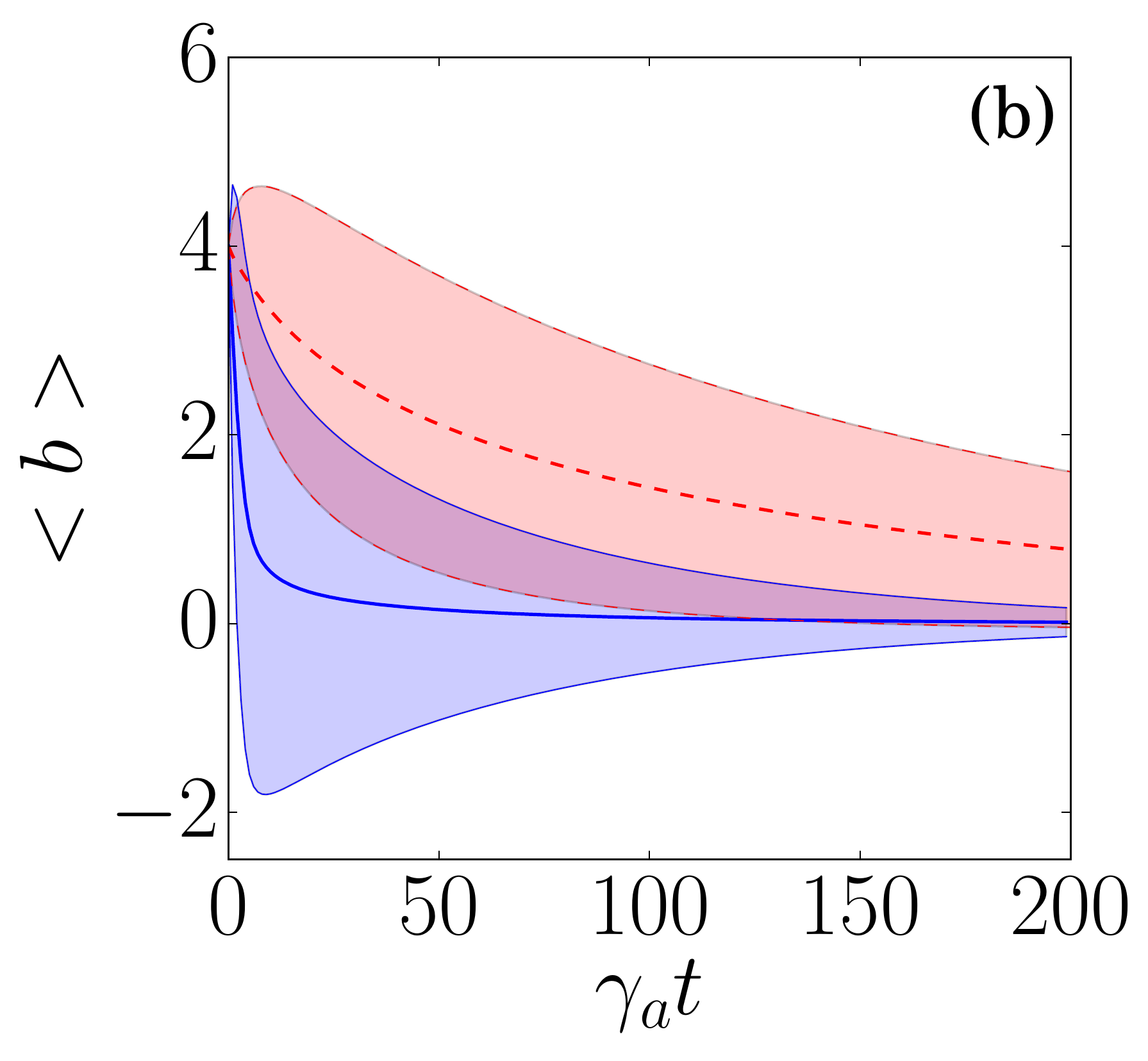}}
\caption{Full master equation prediction for the time evolution of the mirror oscillation amplitude: average value (solid line) and uncertainty due to quantum fluctuations (shaded area). The mirror oscillation is initially prepared in a coherent state of amplitude $\langle \hat{b} \rangle = 4$ and the cavity is initially in its ground state. The two panels refer to the resonant $\Delta/\gamma_a=0$ (a) and non-resonant $\Delta/\gamma_a=10$ (b) cases. In each panel, the two curves refer to the $\gamma_b^{\rm eff}/\gamma_a=1/36$ (blue solid) and $\gamma_b^{\rm eff}/\gamma_a=1/100$ (red dashed) cases.
}
\label{Fig:Shaded}
\end{figure}

Whereas \eqref{Eq:EffGammaB} recovers the predictions of~\cite{Butera-BR_DCE-2019} in the linear limit of small amplitudes $b$,
a remarkable new feature is the strong nonlinearity of the equation of motion for $b$, which is responsible for the peculiar features of the DCE friction force as compared to standard quantum decay~\cite{Petruccione-book,Milburn-Walls}. This effect is well visible in the upper plots of Fig.\ref{Fig:NonLinDecay}(a-c) for $\Delta=0$, in particular in the curves for stronger $\omega_c$ values: while at late times the decay recovers the linear rate $\gamma_b^{\rm eff}$, at early times the rate is significant faster due to the significant amount of DCE photons that are present in the cavity mode $n_a>1$ and stimulate the friction according to the $n_a$-dependence of $\Gamma_b$ predicted by Eq.\ref{Eq:EffGammaB}. The crossover between the stimulated to spontaneous decay occurs once the amplitude $b$ has reached a low enough value such that $n_a<1$.  

Eq.~\eqref{Eq:EffGammaB} naturally provides the validity condition of the BO approximation, which requires that the (slowly varying) amplitude of the mirror oscillation has to evolve on a much longer time scale than the cavity field, which imposes $|\Gamma_b(n_{a,SS})|/\gamma_a\ll 1$ for all values of $n_{a,SS}$ spanned during the evolution. At late times, it imposes that $\omega_c^2/|\gamma_a-i\Delta|\ll\gamma_a$. For macroscopic initial excitations, the bound at short times also involves the initial amplitude $b$ of the oscillator. A quantitatively comparison of the BO prediction with the full Master equation Eq.\ref{Eq:MasterEq} is displayed in Fig.~\ref{Fig:NonLinDecay}(d--f): as expected, the agreement gets worst for higher $\Gamma_b(n_a)/\gamma_a$: for a given initial amplitude $b$, this happens for stronger couplings $\omega_c$, the deviation being particularly visible at early times when the cavity mode is significantly populated by DCE photons.

Whereas the results of the full master equation \eqref{Eq:MasterEq} for the resonant $\Delta=0$ case shown in the upper plots of the panels in Fig.\ref{Fig:NonLinDecay}(a-c) validate the semiclassical theory of backreaction, the situation is more intriguing in the non-resonant cases shown in the lower plots of the panels in Fig.\ref{Fig:NonLinDecay}(a-c): while the intensity $\langle \hat{b}^\dagger \hat{b}\rangle$ of the mirror oscillation keeps decaying at a rate close (modulo the nonlinearity) to the approximate value $\gamma_b^{\rm eff}$, a dramatically reinfored decay is visible for the amplitude $\langle \hat{b}\rangle$. In order to understand this dramatic failure of the semiclassical approximation, we must extend our study to the quantum statistics of the mirror oscillation.

\begin{figure*}[htbp]
\centering
\includegraphics[width=0.95\linewidth]{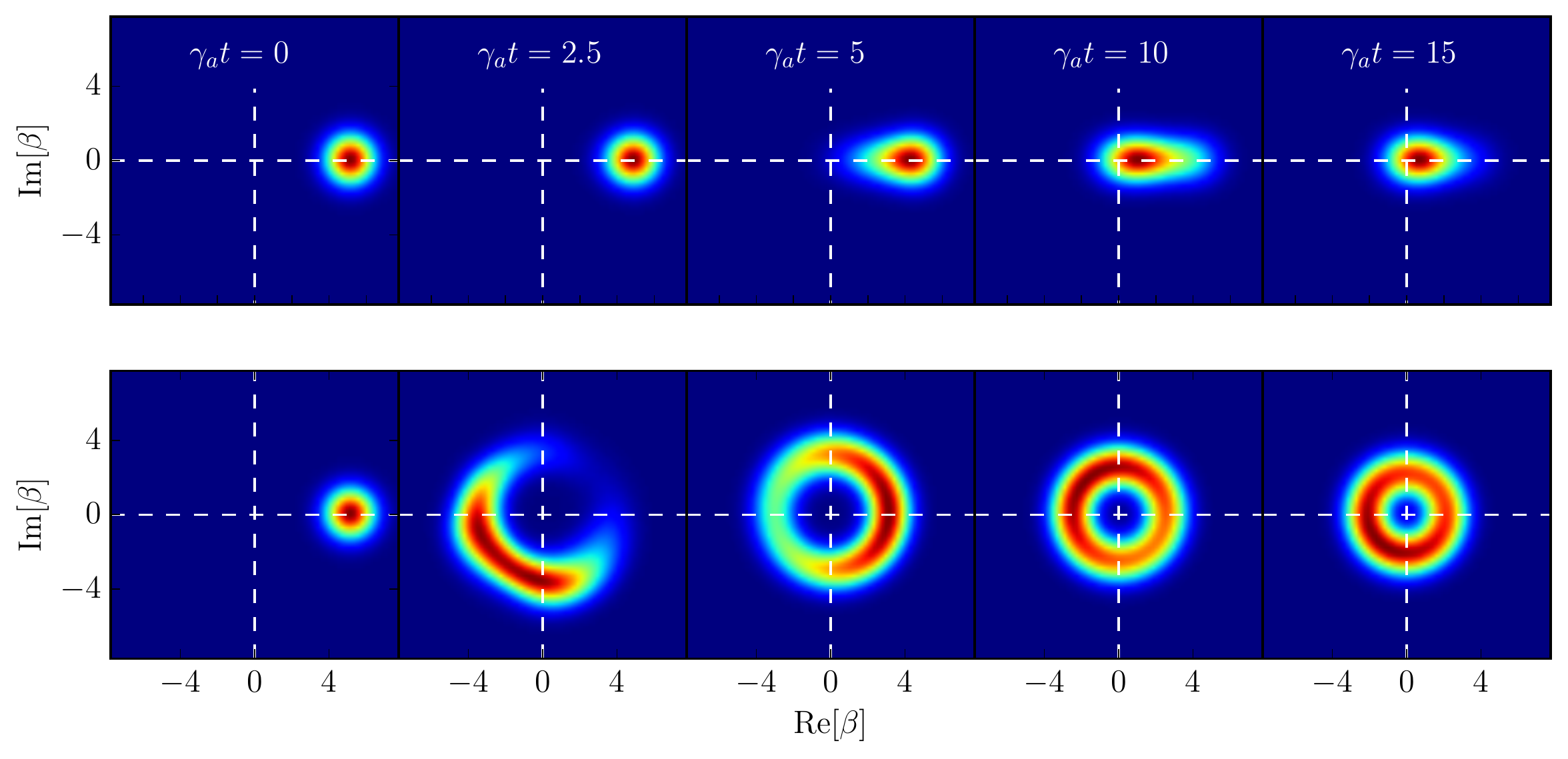}
\caption{Full master equation prediction for the temporal evolution of the $Q$-function of the mechanical oscillator. (a) Resonant case $\Delta/\gamma_a=0$, (b) non-resonant case $\Delta/\gamma_a=10$.}
\label{Fig:Qdens}
\end{figure*}

\begin{figure}[htbp]
\centering
\subfigure
{\includegraphics[width=0.5\linewidth]{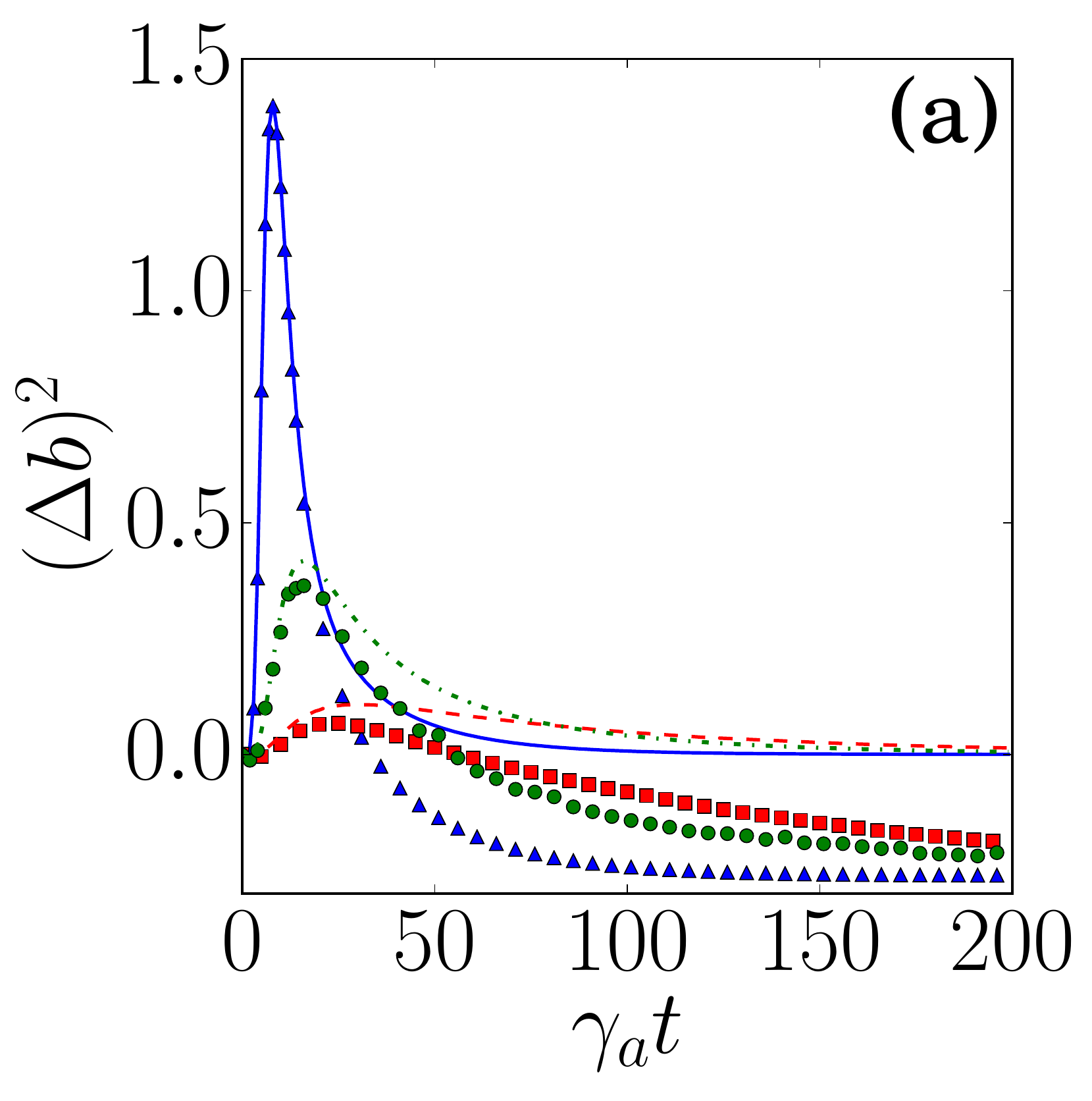}}\hfill
\subfigure
{\includegraphics[width=0.5\linewidth]{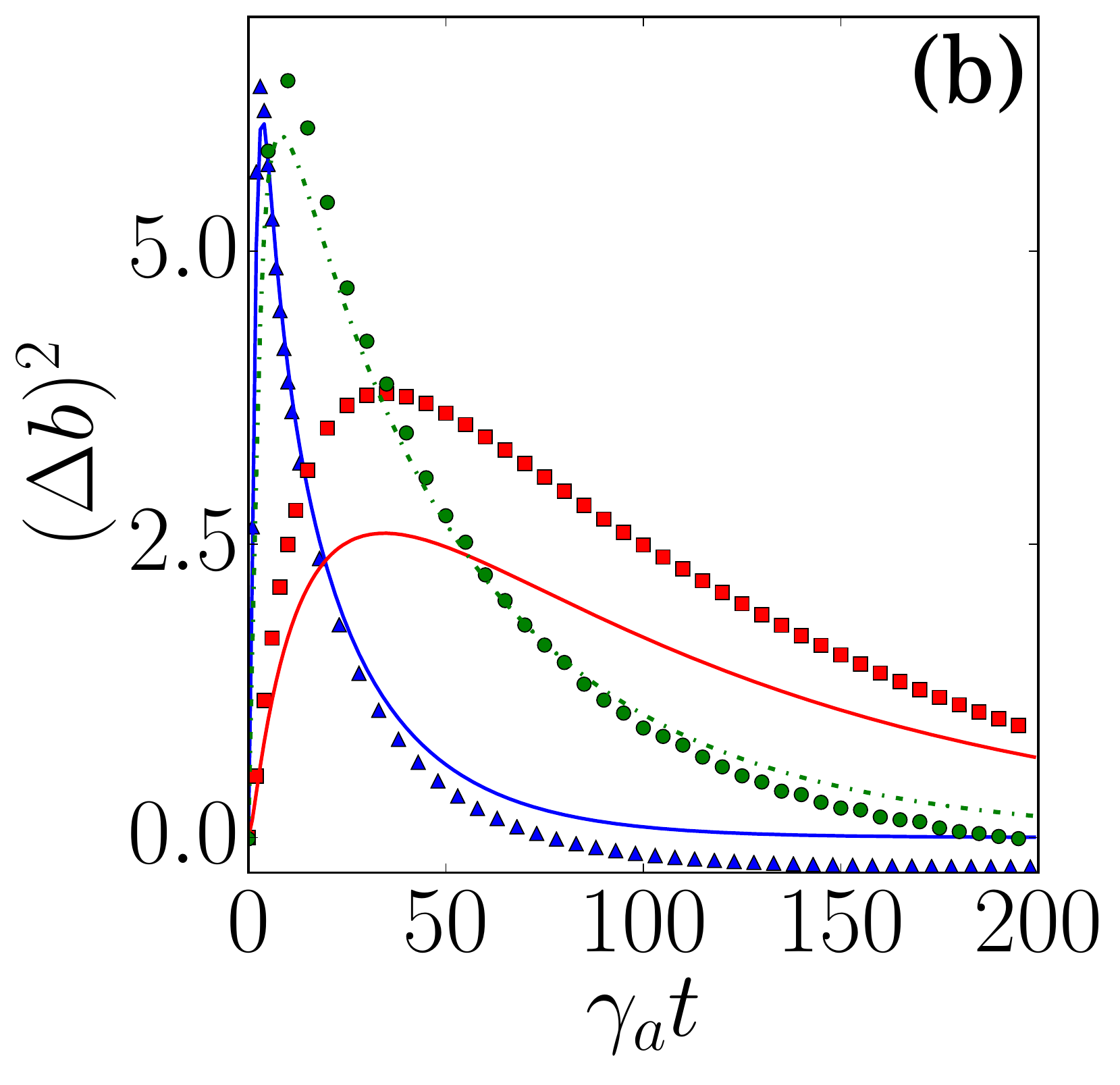}}
\caption{Prediction of the full master equation (solid lines) and of the truncated Wigner approximation (markers) for the time evolution of the variance $\Delta b^2$ of the mirror oscillation amplitude in the resonant $\Delta/\gamma_a=0$ (a) and non-resonant $\Delta/\gamma_a=10$ (b) cases for coupling strength $\omega_c/\gamma_a=1/10$ (blue), $\omega_c/\gamma_a=1/15$ (green) and $\omega_c/\gamma_a=1/20$ (red).}
\label{Fig:TWA}
\end{figure}

\section{Quantum fluctuations of the friction force}
Attacking this problem in terms of the equations of motion Eqs.(\ref{Eq:EqMot-b}-\ref{Eq:EqMot-q}) would require to include higher order correlators and find a reliable way to truncate the resulting infinite hierarchy of equations~\cite{Regemortel-PRA-2017,Qin-Arxiv-2019}. Instead of this, our approach will be again to numerically solve the master equation \eqref{Eq:MasterEq} and then interpret the results in terms of the truncated Wigner approximation (TWA)~\cite{Polkovnikov-AOP-Wigner-2010,Steel-PRA-Wigner-1998,Sinatra-PRA-Wigner-2001}.

Examples of such calculation are shown in Fig.~\ref{Fig:Shaded}(a,b), where we summarize the temporal evolution of the mechanical oscillator amplitude $\hat{b}$ and of its uncertainty 
\begin{equation}
\Delta b^2=\langle \hat{b}^\dagger \hat{b}\rangle -|\langle \hat{b}\rangle|^2
\end{equation}
respectively in the resonant $\Delta=0$ and off-resonant $\Delta/\gamma_a=10$ cases for the same values of $\gamma_b^{\rm eff}$. The average $b=\langle \hat{b}\rangle$ decays along the lines of our previous discussion, with a reinforced decay rate in the non-resonant case. The value of the uncertainty $\Delta b$, indicated by the shading, is also strongly reinforced in this latter case, which confirms that some additional effect must be taking place.


In order to unravel the physical origin of these behaviours, in Fig.\ref{Fig:Qdens} we display the time-evolution of the $Q$-function for the mirror oscillation amplitude, defined as usual as~\cite{Milburn-Walls} 
\begin{equation}
 Q(\beta)=\textrm{Tr}\left[ |\textrm{coh}:\beta \rangle \langle \textrm{coh}:\beta | \,\hat{\rho}\right],
\end{equation}
where $|\textrm{coh}:\beta\rangle$ indicates the coherent state of amplitude $\beta$ and $\hat{\rho}$ is the density matrix of the system. In the resonant case (upper row), the mirror amplitude monotonically decays in time to zero, while the Gaussian-like-shaped $Q$-function is just a bit broadened and stretched along the amplitude direction as compared to the symmetric Gaussian shape of coherent states.

As expected, the situation is completely different in the non-resonant case (bottom row), where the decay is associated to a dramatic reshaping of the distribution. The diffusion of the phase of $\beta$ is in fact very strong and occurs on a time-scale much shorter than the decay of its magnitude $|\beta|$, which results in the ring-like shape of the $Q$-function that is visible in the right-most panels. While the ring square radius $|\beta|^2$ (which physically corresponds to occupation number $\langle \hat{b}^\dagger \hat{b}\rangle$ of the mechanical oscillator) eventually decays on a time scale set by $\gamma_b^{\rm eff}$, the average value of the amplitude $\langle \hat{b}\rangle$ is already very small as soon as the phase has been randomized. This analysis of the phase diffusion explains the marked discrepancy from the semiclassical theory found in Fig.\ref{Fig:NonLinDecay}, in the non-resonant case.

The origin of the reinforced phase diffusion that is observed in the non-resonant case can be intuitively understood by looking back at the (approximate) analytical expression for the effective decay rate $\Gamma_b(n_a)$ given in Eq.\eqref{Eq:EffGammaB}. In the $\Delta=0$ resonant case, the reactive part $\textrm{Im}[\Gamma_b(n_a)]$ vanishes and any fluctuation in the friction force can only result in an elongation of the $Q$-function in the (horizontal) amplitude direction, as it is visible in the central panels of the upper row of Fig.\ref{Fig:Qdens}. On the other hand, in the non-resonant case at finite $\Delta/\gamma_a$, the imaginary part responsible for the reactive frequency shift dominates over the real part responsible for the decay. Given the $n_a$-dependence of $\Gamma_b$, any quantum fluctuation in $n_a$ will then show up as a phase diffusion. 

These physical arguments suggest the possibility of an alternative description of the dynamics in terms of the joint Wigner distribution $W(\alpha,\beta)$ for the cavity field and the mechanical oscillator amplitudes~\cite{Milburn-Walls}. 
Within this formalism, a \emph{classical} limit is naturally identified by the scale transformation $\omega_c\rightarrow\epsilon\omega_c$, $\alpha\rightarrow\alpha/\epsilon$, $\beta\rightarrow\beta/\epsilon$, with $\epsilon\rightarrow 0$, in which the average values of the cavity field and the oscillator amplitude take a macroscopic value, while the nonlinear coupling $\omega_c$ goes to zero. In this limit, third-order derivative terms in the pseudo-Fokker-Planck for the $W$ are 
of higher order in the infinitesimal quantity $\epsilon$ and can thus be neglected according to the so-called {\em truncated Wigner approximation} (TWA)~\cite{Steel-PRA-Wigner-1998,Carusotto2013RMP}. Under this approximation, the evolution of the Wigner distribution $W$ has a Fokker-Planck form and can be recasted in terms of a pair of coupled \^{I}to stochastic differential equations,
\begin{align}
	{dA}&= 
	\left[-i\left(\omega_a-i\frac{\gamma_a}{2}\right)A-2i\omega_c B A^*\right]dt+
 	dW,\label{Eq:SDE-A}\\
	{dB}&=\left[-i\omega_b B-i\omega_c A^2\right]dt \label{Eq:SDE-B}
\end{align}
for the amplitudes of the cavity field $A(t)$ and of the mirror oscillation $B(t)$, where $dW(t)$ is a temporally delta-correlated, zero-average, random phase Gaussian noise such that
\begin{equation}
	\avg{dW^*(t)\,dW(t')}=\frac{\gamma_a}{2}\,\delta(t-t')\,dt\,.
\label{Eq:dB_Corr}
\end{equation}

Conversely to the Born-Oppenheimer approach, that is only reliable for slow decays and imposes an upper bound to the number of photons in the cavity, the TWA description is accurate in the opposite limit where a large number of photons are present in the cavity and the fields behave classically. This is visible in the plots of the average amplitude shown in Fig.~\ref{Fig:NonLinDecay}(d--f) as well as in the ones of the variance $\Delta b^2$ shown in Fig.~\ref{Fig:TWA}. In this latter figure, the TWA prediction (markers) is compared to the master equation Eq.~\eqref{Eq:MasterEq} for a given initial amplitude $b_0=4$ and different coupling strengths $\omega_c/\gamma_a=1/20$ (blue curves), $1/15$ (green curves) and $1/10$ (red curved) in both the resonant $\Delta=0$ [panel (a)] and non-resonant $\Delta/\gamma_a=10$ [panel(b)] cases. As expected, the agreement is here very good at short times where the occupation of both modes are large and, quite remarkably, this condition is better satisfied for stronger values of the coupling $\omega_c$. On the other hand, the discrepancies observed at late times (including the non-physical negative occupations) are due to the small occupation and to the well-known pathologies of the truncated Wigner method in this regime. 

From a physical point of view, the accuracy of the reformulation in terms of the stochastic TWA equations confirms our interpretation of the numerical master equation results in terms of phase diffusion: the quantum fluctuations due to the stochastic noise term $dW$ are responsible for a wide distribution of the cavity field amplitude $A(t)$ around its average value. Because of the nonlinear form of the motion equations (\ref{Eq:SDE-A}-\ref{Eq:SDE-B}), this results in a fluctuating frequency of the mirror oscillations analogous to \eqref{Eq:EffGammaB}.

\section{Theoretical considerations}
To place our results into a wider context, it is interesting to draw a connection between our TWA approach and the so-called stochastic gravity framwork~\cite{Hu-StocGrav-CGQ,Hu-StocGrav-LivRev}. The equation of motion of the TWA have the form of stochastic differential equations for the semiclassical amplitudes and provide a relation between the noise in the quantum field, in our case the cavity field, and the fluctuations in the background, whose unique degree of freedom is identified in our DCE case with the motion of the mirror. Within its limits of validity, the TWA is an accurate approximation of the full quantum dynamics of the system as described by the master equation in Eq.~\eqref{Eq:MasterEq}. As such, it can be used to completely reconstruct the full set of correlators for the mechanical oscillator, which is the equivalent of the so called Boltzmann--Einstein hierarchy of quantum gravity \cite{Hu2002}.

\section{Experimental considerations}
It is well-known that experimental studies of the dynamical Casimir effect are made extremely challenging by the extremely weak intensity of the emission~\cite{Lambrecht-2005}. To overcome this problem and get access to the basic physics of quantum field theories in modulated and curved space-times, a steadily growing community has started looking at the so-called analog models~\cite{Unruh-Analog-1981,Volovik2001,Barcelo-2011,faccio-book}. These consist of condensed matter systems displaying some excitation mode that, in suitable limits, can be described in terms of a quantum field theory on a curved and/or modulated space-time. Intense theoretical efforts on atomic Bose-Einstein condensates~\cite{Garay2000,Leonhardt2003b,Fedichev2004,Uhlmann2005,Fischer2004,jain2007,Carusotto2008,IC_EPJD,Jaskula2012,finazzi2014} and superconducting circuits~\cite{Johansson-PRL-2009,Schutzhold2005,Nation-PRL-2009,Nation-RMP-2012} have paved the way to recent experimental evidences of (analog) dynamical Casimir~\cite{Wilson-DCE-Analog-2011,Lahteenmaki-DCE-Analog-2013} and Hawking emissions~\cite{steinhauer2016observation}. The next challenge will be to push the research on analog models forward towards the backreaction problem: pioneering theoretical developments have focussed on the Hawking emission~\cite{Balbinot-BR_PRL-2005,Balbinot-BR_PRD-2005,Maia-BR-2007,Maia-BR-2008,Plunien-BR-2007,nation2010trilinear} and the DCE~\cite{Carusotto-PRA-AnBackR-2012} cases. In order to implement the specific proposal of this Letter, promising candidates are the superconducting circuit devices used in~\cite{Wilson-DCE-Analog-2011,Lahteenmaki-DCE-Analog-2013}, whose potential to observe backreaction effects was discussed in~\cite{Savasta-PRX-2018,Butera-BR_DCE-2019}. 

\section{Conclusions}
In this work we have reported a theoretical study of the quantum fluctuations of the friction force exerted by the dynamical Casimir emission onto a moving mirror. Capitalizing on our previous work on the average friction force~\cite{Butera-BR_DCE-2019}, observable signatures of the quantum fluctuations have been identified in the quantum state of the mechanical oscillation of the mirror, which turns out to dramatically depart from its initial coherent form. Even though our work is based on a simplest, yet realistic model of dynamical Casimir emission in an optical cavity that is amenable to exact calculations at a fully quantum level, we expect that our findings on the dramatic breakdown of the semiclassical approximation and the non-trivial quantum statistics of the mechanical oscillation are very general and may have deep consequences for a variety of problems in quantum field theories on curved spacetimes and in gravitation. Next steps will include an investigation of the conceptual links between our findings and stochastic gravity models~\cite{Hu-StocGrav-LivRev} and the extension of our study of backreaction effects to those multi-mode configurations that naturally appear in dynamical Casimir experiments with atomic fluids~\cite{Carusotto2008,Jaskula2012,eckel2018rapidly,Robertson-PRD-PreHeatAn-2019}. These investigations will pave the way towards the more challenging task of understanding quantum fluctuation features in the black hole evaporation process~\cite{fabbri2005modeling}.

\section{Acknowledgements}
Stimulating discussions with R. Balbinot, R. Parentani, B. L. Hu and M. Rinaldi are warmly acknowledged. This work was supported by the Julian  Schwinger foundation, Grant No. JSF-16-12-0001. I.C. acknowledges funding from  Provincia Autonoma di Trento and from the Quantum Flagship Grant PhoQuS (820392) of the European Union.

\bibliography{DCE-BR.bib}

\begin{thebibliography}{10}
\expandafter\ifx\csname url\endcsname\relax\def\url#1{\texttt{#1}}\fi

\bibitem{Milonni-book}
\Name{Milonni P.~W.} \Book{The quantum vacuum: an introduction to quantum
  electrodynamics} (Academic press New York) 1994.

\bibitem{birrell1984quantum}
\Name{Birrell N.~D. \and Davies P. C.~W.} \Book{Quantum Fields in Curved Space}
  Cambridge Monographs on Mathematical Physics (Cambridge University Press)
  1984.

\bibitem{Moore-DCE-1970}
\Name{Moore G.~T.} \REVIEW{J. Math. Phys.}{11}{1970}{2679}.

\bibitem{Fulling_Davies-DCE}
\Name{Fulling S.~A. \and Davies P. C.~W.} \REVIEW{Proc. R. Soc. Lond. A Math.
  Phys. Sci.}{348}{1976}{393}.

\bibitem{Davies-Fulling}
\Name{Davies P. C.~W. \and Fulling S.~A.} \REVIEW{Proc. R. Soc. Lond. A Math.
  Phys. Sci.}{356}{1977}{237}.

\bibitem{Dodonov-DCE}
\Name{Dodonov V.~V.} \REVIEW{Physica Scripta}{82}{2010}{038105}.

\bibitem{lambrecht2005electromagnetic}
\Name{Lambrecht A.} \REVIEW{Journal of Optics B: Quantum and Semiclassical
  Optics}{7}{2005}{S3}.

\bibitem{KardarRMP1999}
\Name{Kardar M. \and Golestanian R.} \REVIEW{Rev. Mod. Phys.}{71}{1999}{1233}.

\bibitem{Parker-PartCr-I}
\Name{Parker L.} \REVIEW{Phys. Rev.}{183}{1969}{1057}.

\bibitem{Parker-PartCr-II}
\Name{Parker L.} \REVIEW{Phys. Rev. D}{3}{1971}{346}.

\bibitem{Hawking1974}
\Name{Hawking S.~W.} \REVIEW{Nature}{248}{1974}{30}.

\bibitem{Hawking1975}
\Name{Hawking S.~W.} \REVIEW{Commun. Math. Phys}{43}{1975}{199}.

\bibitem{fabbri2005modeling}
\Name{Fabbri A. \and Navarro-Salas J.} \Book{Modeling Black Hole Evaporation}
  (Imperial College Press) 2005.

\bibitem{Hu-StocGrav-CGQ}
\Name{Hu B.-L. \and Verdaguer E.} \REVIEW{Class. Quantum Grav.}{20}{2003}{R1}.

\bibitem{Hu-StocGrav-LivRev}
\Name{Hu B.-L. \and Verdaguer E.} \REVIEW{Living Rev. Relativ.}{11}{2008}{3}.

\bibitem{Butera-BR_DCE-2019}
\Name{Butera S. \and Carusotto I.} \REVIEW{Phys. Rev. A}{99}{2019}{053815}.

\bibitem{Petruccione-book}
\Name{Breuer H.~P., Petruccione F. \etal} \Book{The theory of open quantum
  systems} (Oxford University Press) 2002.

\bibitem{Milburn-Walls}
\Name{Walls D.~F. \and Milburn G.~J.} \Book{{Quantum Optics}} (Springer) 2008.

\bibitem{Aspelmeyer_RMP}
\Name{Aspelmeyer M., Kippenberg T.~J. \and Marquardt F.} \REVIEW{Rev. Mod.
  Phys.}{86}{2014}{1391}.

\bibitem{Law-MirFieldInt-1995}
\Name{Law C.~K.} \REVIEW{Phys. Rev. A}{51}{1995}{2537}.

\bibitem{CohenTannoudji-AtomPhot}
\Name{Cohen-Tannoudji C., Dupont-Roc J. \and Grynberg G.} \Book{{Atom-Photon
  Interactions: Basic Processes and Applications}} (Wiley-VCH) 1998.

\bibitem{Savasta-PRX-2018}
\Name{Macr\`{\i} V., Ridolfo A., Di~Stefano O., Kockum A.~F., Nori F. \and
  Savasta S.} \REVIEW{Phys. Rev. X}{8}{2018}{011031}.

\bibitem{Regemortel-PRA-2017}
\Name{Van~Regemortel M., Casteels W., Carusotto I. \and Wouters M.}
  \REVIEW{Phys. Rev. A}{96}{2017}{053854}.

\bibitem{Qin-Arxiv-2019}
\Name{Qin W., Macr\`{i} V., Miranowicz A., Savasta S. \and Nori F.}
  \REVIEW{arXiv:1902.04216}{}{2019}{}.

\bibitem{Polkovnikov-AOP-Wigner-2010}
\Name{Polkovnikov A.} \REVIEW{Ann. Phys.}{325}{2010}{1790 }.

\bibitem{Steel-PRA-Wigner-1998}
\Name{Steel M.~J., Olsen M.~K., Plimak L.~I., Drummond P.~D., Tan S.~M.,
  Collett M.~J., Walls D.~F. \and Graham R.} \REVIEW{Phys. Rev.
  A}{58}{1998}{4824}.

\bibitem{Sinatra-PRA-Wigner-2001}
\Name{Sinatra A., Lobo C. \and Castin Y.} \REVIEW{Phys. Rev.
  Lett.}{87}{2001}{210404}.

\bibitem{Carusotto2013RMP}
\Name{Carusotto I. \and Ciuti C.} \REVIEW{Rev. Mod. Phys.}{85}{2013}{299}.

\bibitem{Hu2002}
\Name{Hu B.~L.} \REVIEW{‎Int. J. Theor. Phys.}{41}{2002}{2091}.

\bibitem{Lambrecht-2005}
\Name{Lambrecht A.} \REVIEW{Journal of Optics B: Quantum and Semiclassical
  Optics}{7}{2005}{S3}.

\bibitem{Unruh-Analog-1981}
\Name{Unruh W.~G.} \REVIEW{Phys. Rev. Lett.}{46}{1981}{1351}.

\bibitem{Volovik2001}
\Name{Volovik G.~E.} \REVIEW{‎Phys. Rep.}{351}{2001}{195 }.

\bibitem{Barcelo-2011}
\Name{Barcel{\'o} C., Liberati S. \and Visser M.} \REVIEW{Living Rev.
  Relativ.}{14}{2011}{}.

\bibitem{faccio-book}
\Name{Faccio D., Belgiorno F., Cacciatori S., Gorini V., Liberati S. \and
  Moschella U.} \REVIEW{}{}{}{}.

\bibitem{Garay2000}
\Name{Garay L.~J., Anglin J.~R., Cirac J.~I. \and Zoller P.} \REVIEW{Phys. Rev.
  Lett.}{85}{2000}{4643}.

\bibitem{Leonhardt2003b}
\Name{Leonhardt U., Kiss T. \and \"Ohberg P.} \REVIEW{J. Opt. B}{5}{2003}{S42}.

\bibitem{Fedichev2004}
\Name{Fedichev P.~O. \and Fischer U.~R.} \REVIEW{Phys. Rev.
  A}{69}{2004}{033602}.

\bibitem{Uhlmann2005}
\Name{Uhlmann M., Xu Y. \and Sch{\"u}tzhold R.} \REVIEW{New J.
  Phys.}{7}{2005}{248}.

\bibitem{Fischer2004}
\Name{Fischer U.~R. \and Sch\"utzhold R.} \REVIEW{Phys. Rev.
  A}{70}{2004}{063615}.

\bibitem{jain2007}
\Name{Jain P., Weinfurtner S., Visser M. \and Gardiner C.~W.} \REVIEW{Phys.
  Rev. A}{76}{2007}{033616}.

\bibitem{Carusotto2008}
\Name{Carusotto I., Fagnocchi S., Recati A., Balbinot R. \and Fabbri A.}
  \REVIEW{New J. Phys.}{10}{2008}{103001}.

\bibitem{IC_EPJD}
\Name{Carusotto I., Balbinot R., Fabbri A. \and Recati A.} \REVIEW{Eur. Phys.
  J. D}{56}{2010}{391}.

\bibitem{Jaskula2012}
\Name{Jaskula J.-C., Partridge G.~B., Bonneau M., Lopes R., Ruaudel J., Boiron
  D. \and Westbrook C.~I.} \REVIEW{Phys. Rev. Lett.}{109}{2012}{220401}.

\bibitem{finazzi2014}
\Name{Finazzi S. \and Carusotto I.} \REVIEW{Phys. Rev. A}{90}{2014}{033607}.

\bibitem{Johansson-PRL-2009}
\Name{Johansson J.~R., Johansson G., Wilson C.~M. \and Nori F.} \REVIEW{Phys.
  Rev. Lett.}{103}{2009}{147003}.

\bibitem{Schutzhold2005}
\Name{Sch\"utzhold R. \and Unruh W.~G.} \REVIEW{Phys. Rev.
  Lett.}{95}{2005}{031301}.

\bibitem{Nation-PRL-2009}
\Name{Nation P.~D., Blencowe M.~P., Rimberg A.~J. \and Buks E.} \REVIEW{Phys.
  Rev. Lett.}{103}{2009}{087004}.

\bibitem{Nation-RMP-2012}
\Name{Nation P.~D., Johansson J.~R., Blencowe M.~P. \and Nori F.} \REVIEW{Rev.
  Mod. Phys.}{84}{2012}{1}.

\bibitem{Wilson-DCE-Analog-2011}
\Name{Wilson C.~W., Johansson G., Pourkabirian A., Simoen M., Johansson J.~R.,
  Duty T., Nori F. \and Delsing P.} \REVIEW{Nature}{479}{2011}{376}.

\bibitem{Lahteenmaki-DCE-Analog-2013}
\Name{L{\"a}hteenm{\"a}ki P., Paraoanu G.~S., Hassel J. \and Hakonen P.~J.}
  \REVIEW{Proc. Natl. Acad. Sci.}{110}{2013}{4234}.

\bibitem{steinhauer2016observation}
\Name{Steinhauer J.} \REVIEW{Nature Physics}{12}{2016}{959}.

\bibitem{Balbinot-BR_PRL-2005}
\Name{Balbinot R., Fagnocchi S., Fabbri A. \and Procopio G.~P.} \REVIEW{Phys.
  Rev. Lett.}{94}{2005}{161302}.

\bibitem{Balbinot-BR_PRD-2005}
\Name{Balbinot R., Fagnocchi S. \and Fabbri A.} \REVIEW{Phys. Rev.
  D}{71}{2005}{064019}.

\bibitem{Maia-BR-2007}
\Name{Maia C. \and Sch\"utzhold R.} \REVIEW{Phys. Rev. D}{76}{2007}{101502}.

\bibitem{Maia-BR-2008}
\Name{Sch\"utzhold R. \and Maia C.} \REVIEW{J. Phys. A: Math.
  Theor.}{41}{2008}{164065}.

\bibitem{Plunien-BR-2007}
\Name{Plunien G., Ruser M. \and Sch\"utzhold R.} \REVIEW{Class. Quantum
  Grav.}{24}{2008}{4361}.

\bibitem{nation2010trilinear}
\Name{Nation P.~D. \and Blencowe M.~P.} \REVIEW{New Journal of
  Physics}{12}{2010}{095013}.

\bibitem{Carusotto-PRA-AnBackR-2012}
\Name{Carusotto I., De~Liberato S., Gerace D. \and Ciuti C.} \REVIEW{Phys. Rev.
  A}{85}{2012}{023805}.

\bibitem{eckel2018rapidly}
\Name{Eckel S., Kumar A., Jacobson T., Spielman I.~B. \and Campbell G.~K.}
  \REVIEW{Physical Review X}{8}{2018}{021021}.

\bibitem{Robertson-PRD-PreHeatAn-2019}
\Name{Robertson S., Michel F. \and Parentani R.} \REVIEW{Phys. Rev.
  D}{98}{2018}{056003}.

\end{thebibliography}
\bibliographystyle{eplbib}

\end{document}